\documentclass[preprint,aps, pra,amsmath,amssymb, floatfix]{revtex4-2}

\usepackage{filecontents}

\usepackage{graphicx}
\usepackage{dcolumn}
\usepackage{bm}
\usepackage[utf8]{inputenc}
\usepackage[T1]{fontenc}
\usepackage{mathptmx}
\usepackage{etoolbox}
\usepackage{xcolor}
\usepackage{hyperref}

\renewcommand{\selectlanguage}[1]{}

\makeatletter
\def\@email#1#2{%
 \endgroup
 \patchcmd{\titleblock@produce}
  {\frontmatter@RRAPformat}
  {\frontmatter@RRAPformat{\produce@RRAP{*#1\href{mailto:#2}{#2}}}\frontmatter@RRAPformat}
  {}{}
}%
\makeatother

\begin{document}

\title{Collecting single photons from a cavity-coupled quantum dot using an adiabatic tapered fiber}
\author{A. Bach}
\altaffiliation{anatole.bach@webmail.insp.jussieu.fr}
\affiliation{Sorbonne Université, CNRS, Institut des Nanosciences de Paris, 75005 Paris, France}
\author{A. Chapuis}
\affiliation{Sorbonne Université, CNRS, Institut des Nanosciences de Paris, 75005 Paris, France}
\author{K. Moratis}
\affiliation{Université Paris-Saclay, CNRS, Centre de Nanosciences et de Nanotechnologies, 91120, Palaiseau, France}
\author{C. Morin}
\affiliation{Sorbonne Université, CNRS, Institut des Nanosciences de Paris, 75005 Paris, France}
\author{R. Hostein}
\affiliation{Sorbonne Université, CNRS, Institut des Nanosciences de Paris, 75005 Paris, France}
\author{S. Germanis}
\affiliation{Sorbonne Université, CNRS, Institut des Nanosciences de Paris, 75005 Paris, France}
\author{B. Eble}
\affiliation{Sorbonne Université, CNRS, Institut des Nanosciences de Paris, 75005 Paris, France}
\author{M. Bernard}
\affiliation{Sorbonne Université, CNRS, Institut des Nanosciences de Paris, 75005 Paris, France}
\author{F. Margaillan}
\affiliation{Sorbonne Université, CNRS, Institut des Nanosciences de Paris, 75005 Paris, France}
\author{R. Braive}
\affiliation{Université Paris-Saclay, CNRS, Centre de Nanosciences et de Nanotechnologies, 91120, Palaiseau, France}
\affiliation{Université Paris Cité, CNRS, Centre de Nanosciences et de Nanotechnologies, 91120 Palaiseau, France}
\affiliation{Institut Universitaire de France (IUF), Paris, France}
\author{P. Atkinson}
\affiliation{Sorbonne Université, CNRS, Institut des Nanosciences de Paris, 75005 Paris, France}
\author{V. Voliotis}
\affiliation{Sorbonne Université, CNRS, Institut des Nanosciences de Paris, 75005 Paris, France}
\date{\today}

\begin{abstract}
We demonstrate efficient in-plane optical fiber collection of single photon emission from quantum dots embedded in photonic crystal cavities. This was achieved via adiabatic coupling between a tapered optical fiber and a tapered on-chip photonic waveguide coupled to the photonic crystal cavity. 
The collection efficiency of a dot in a photonic crystal cavity was measured to be 5 times greater via the tapered optical fiber compared to collection by a microscope objective lens above the cavity. 
The single photon source was also characterized by second order photon correlations measurements giving g$^{(2)}$(0)=0.17 under non-resonant excitation.
Numerical calculations demonstrate that the collection efficiency could be further increased by improving the dot-cavity coupling and by increasing the overlap length of the tapered fiber with the on-chip waveguide. An adiabatic coupling of near unity is predicted for an overlap length of 5 $\mu$m.
\end{abstract}

\maketitle

\section{Introduction}

Self-assembled semiconductor quantum dots (QDs) have demonstrated their great potential for quantum information applications ranging from bright single and indistinguishable photon sources
\cite{coste_high-rate_2023}
to spin qubit operation \cite{michler_quantum_2017}. 

Specific photonic nanostructures with embedded QDs are promising platforms for scalable quantum communications protocols\cite{uppu_scalable_2020,heindel_quantum_2023}. 
Shaping the density of states of the electromagnetic field in the vicinity of the QD can increase the light-matter coupling and enhance the spontaneous emission rate through the Purcell effect\cite{purcell_e_proceedings_1946}. 
This reduction in radiative lifetime not only increases the photon emission rate but it also reduces the impact of dephasing processes, caused by the presence of charge \cite{reigue_resonance_2019} and spin noise \cite{kuhlmann_charge_2013} or due to coupling to the phonon bath \cite{reigue_probing_2017}, resulting in a higher degree of photon indistinguishability \cite{liu_high_2018}.

Photonic crystal (PhC) slabs with embedded cavities or waveguides show many advantages due to their planar geometry. Besides the achieved strong light-matter coupling \cite{hennessy_quantum_2007,englund_resonant_2010}, they can also be linked together or to an input-output optical signal through on-chip waveguides for efficient transfer of photons \cite{arcari_near-unity_2014, liu_high_2018}, making them promising platforms for spin-photon interfaces \cite{arcari_near-unity_2014}. 
Nevertheless, spatial and spectral matching between the emitter and the optical mode remain an issue for scalable architectures.
A way to overcome the problem of spatial matching is to realize photoluminescence (PL) mapping to determine the position of the QDs prior to the etching of the PhC nanostructures at the specific positions. 
This method can be very accurate if specific markers on the sample surface are used combined to a well designed optical setup\cite{sapienza_nanoscale_2015,kojima_accurate_2013,liu_cryogenic_2017}.

Another major challenge still to be met is the efficient extraction of photons from the PhC nanostructures. 
An interesting approach is to realize circular Bragg gratings that can reach 85\%  of on-chip extraction efficiency \cite{destouches_efficient_2008, faraon_dipole_2008}. 
Furthermore, efficient fiber-coupled single photon sources are highly desirable for next generation scalable devices and important progresses have been made recently towards this goal \cite{bremer_fiber-coupled_2022, de_gregorio_plug-and-play_2024}. An interesting approach 
with theoretical extraction efficiency close to 100\% 
is the adiabatic coupling of waveguides with optical fibers. This approach has been explored successfully in the case of silicon or diamond waveguides \cite{khan_low-loss_2020,burek_2017} coupled to tapered fibers and in the case of nanobeam waveguides coupled off-chip to lensed fibers \cite{kirsanke_indistinguishable_2017}.
Collection of single photons emitted by a self-assembled QD coupled to a photonic nanostructure using adiabatic coupling with a tapered fiber has been demonstrated in the case of tens of microns long waveguides with a chip-to-fiber collection efficiency around 80\%, measured by reflection measurements \cite{daveau_efficient_2017}. Yet, collection of single
photons emitted by a QD coupled to a PhC cavity using adiabatic coupling with a
tapered fiber has not been demonstrated to the best of our knowledge.


Here, we study the efficiency of in-plane fiber collection of the emission from an InAs QD embedded in a photonic crystal cavity. The in-plane collection is achieved by first coupling the photonic crystal cavity to a photonic crystal waveguide which then guides the emission into a suspended tapered GaAs waveguide. The emission is then collected via adiabatic coupling by a tapered optical fiber placed in close proximity to the tapered GaAs waveguide. 
In section II the sample design and the experimental setup are presented. Section III presents the optical characterization of the QD-PhC coupled system. 
We demonstrate Purcell enhanced spontaneous emission of a cavity-coupled QD and collection of single photons with a g$^{(2)}$(0)=0.17 through an adiabatically coupled tapered fiber. We discuss the possible losses terms that appear in these kinds of complex nanophotonic architectures and propose paths for future improvements.

\section{Sample design and experimental setup}

The sample is grown by molecular beam epitaxy (MBE) and consists of a self-assembled InAs QDs layer embedded in a 180 nm thick GaAs membrane on top of an AlGaAs sacrificial layer with 500 nm thickness. 
Micro-photoluminescence ($\mu$PL) mapping of the sample using Ti/Au markers (8 $\mu$m crosses separated by 40 $\mu$m) deposited on the surface allows to determine the position of the QDs with respect to the markers with an accuracy better than $\pm$200 nm (see Appendix \ref{app:mapping} for more details). These markers are also used as alignment marks for the fabrication of the photonic nanostructures which are realized using electron beam lithography and ICP etching followed by chemical etching to remove the sacrificial layer and obtain the suspended membrane \cite{braive_inductively_2009, midolo_soft-mask_2015}. 

\begin{figure}
    \centering
    \includegraphics[width=0.45\textwidth]{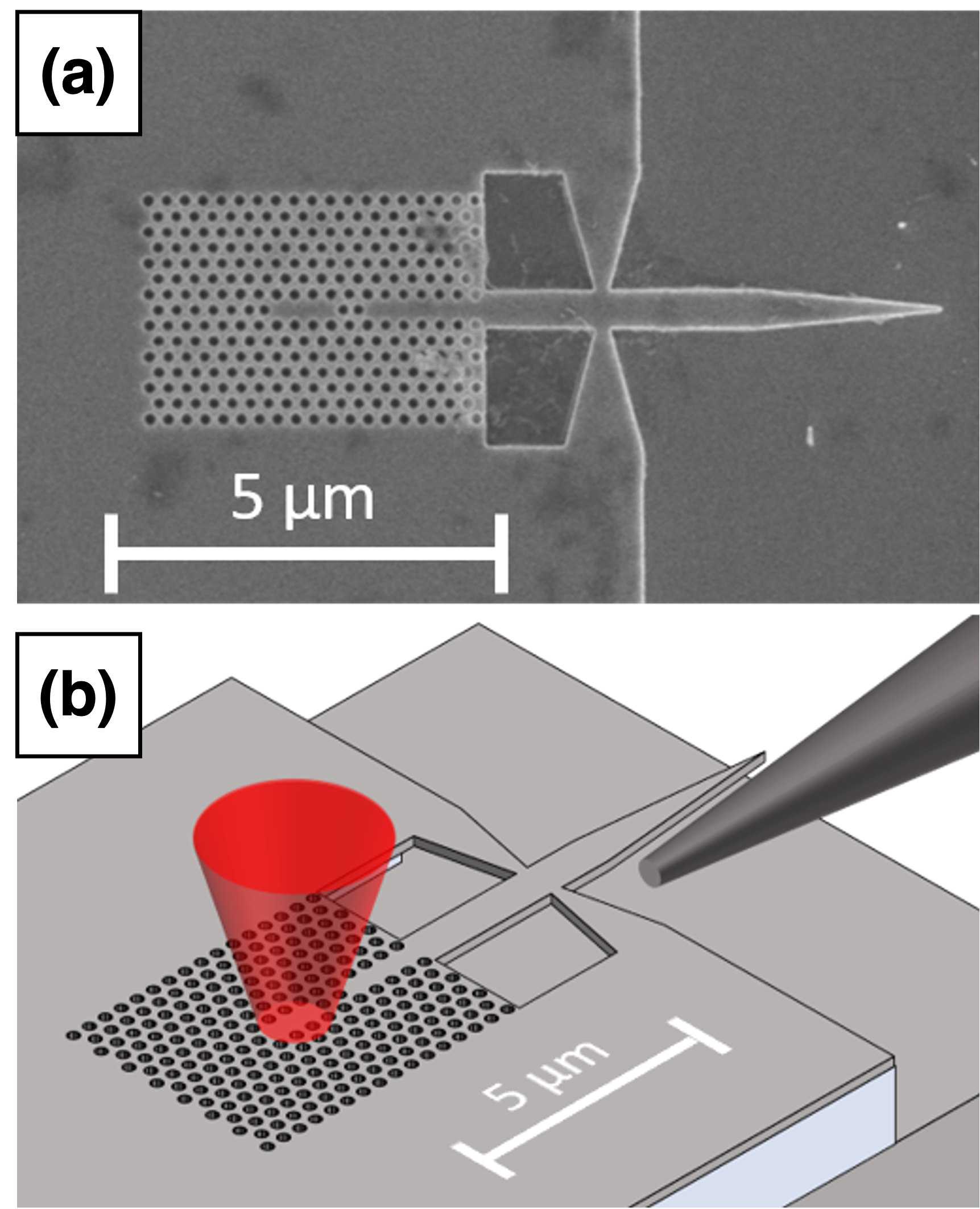}
    \caption{(a) SEM image of the fabricated device. A L3 cavity is coupled to a PhC waveguide, which guides the emitted light into a conventional bulk waveguide. The triangular shape of the bulk waveguide is designed for adiabatic coupling with a tapered fiber. (b) Schematic of the excitation/collection setup: the PhC cavity can be excited by a top microscope objective with NA 0.65 (red cone) and the emitted photons can be collected either by the same objective or by the waveguide-coupled tapered fiber (in dark gray).}
    \label{fig:image_MEB_II}
\end{figure}

A scanning electron microscopy (SEM) image of the photonic nanostructure is presented in Fig. \ref{fig:image_MEB_II}a. 
A L3 PhC cavity is coupled to a PhC waveguide (PCWG) followed by a rectangular GaAs waveguide ending in a triangular shape in the two dimensions. 
The base width of the tapered waveguide is 500 nm and the width decreases progressively over 2500 nm. The waveguide mode is then adiabatically coupled to a tapered fiber, represented in dark grey Fig. \ref{fig:image_MEB_II}b, that can be approached in the vicinity of the waveguide using nanopositioners. 

The photonic nanostructures have been optimized by modifying the hole radii and positions around the cavity \cite{vuckovic_design_2001,vuckovic_photonic_2003} to enhance the coupling of the photonic cavity with the PCWG and the GaAs waveguide, in order to increase the collection of the emitted photons by the tapered fiber. 
A hundred photonic nanostructures have been characterized. The dispersion of the hole radii of the fabricated PhC cavities, as observed on the SEM images, induced an uncertainty of $\pm$ 5 meV in the central energy position of the optical mode of the cavity. Nevertheless, we present here results on a QD whose emission is sufficiently coupled to the cavity to demonstrate enhanced off-chip collection via the tapered fiber.
The relevant parameters of the photonic nanostructure are given in Appendix \ref{app:pcparam}.
The theoretical Purcell factor for a QD perfectly matched spectrally and spatially to the L3 cavity's fundamental mode with quality factor Q, mode volume V$_{mode}$ and central wavelength $\lambda$ is {$F_P =(3/4\pi^2)(Q/V_{mode})(\lambda/n)^3 = 413$}. 

A silica monomode fiber was tapered using a laser based micropipette puller (P-2000 \textit{Sutter Instrument}), resulting in a tip diameter of 350 nm. 
Its tapered end is held in a V-groove holder mounted on a piezoelectric stack, allowing the fiber to be approached close to the waveguide and to tune the size of the "superguide" formed by the waveguide and the fiber, as defined in Fig.\ref{fig:schem_couplage}, with nanometric precision. 
The fiber passes through an opening of a MyCryoFirm Optidry 250 ultralow vibration cryostat. 
It is then fused outside the cryostat with another cleaved optical fiber using a fusion splicer (S179A \textit{FITEL}).
This optical fiber can then be connected to the different detection setups for micro-photoluminescence (µPL) spectroscopy or photon correlations measurements. 
The whole setup is robust to mechanical vibrations so that the optical fiber position is stable for several days.

\section{Results and discussion}
\label{sec:results}

\begin{figure*}
    \centering
    \includegraphics[width=\textwidth]{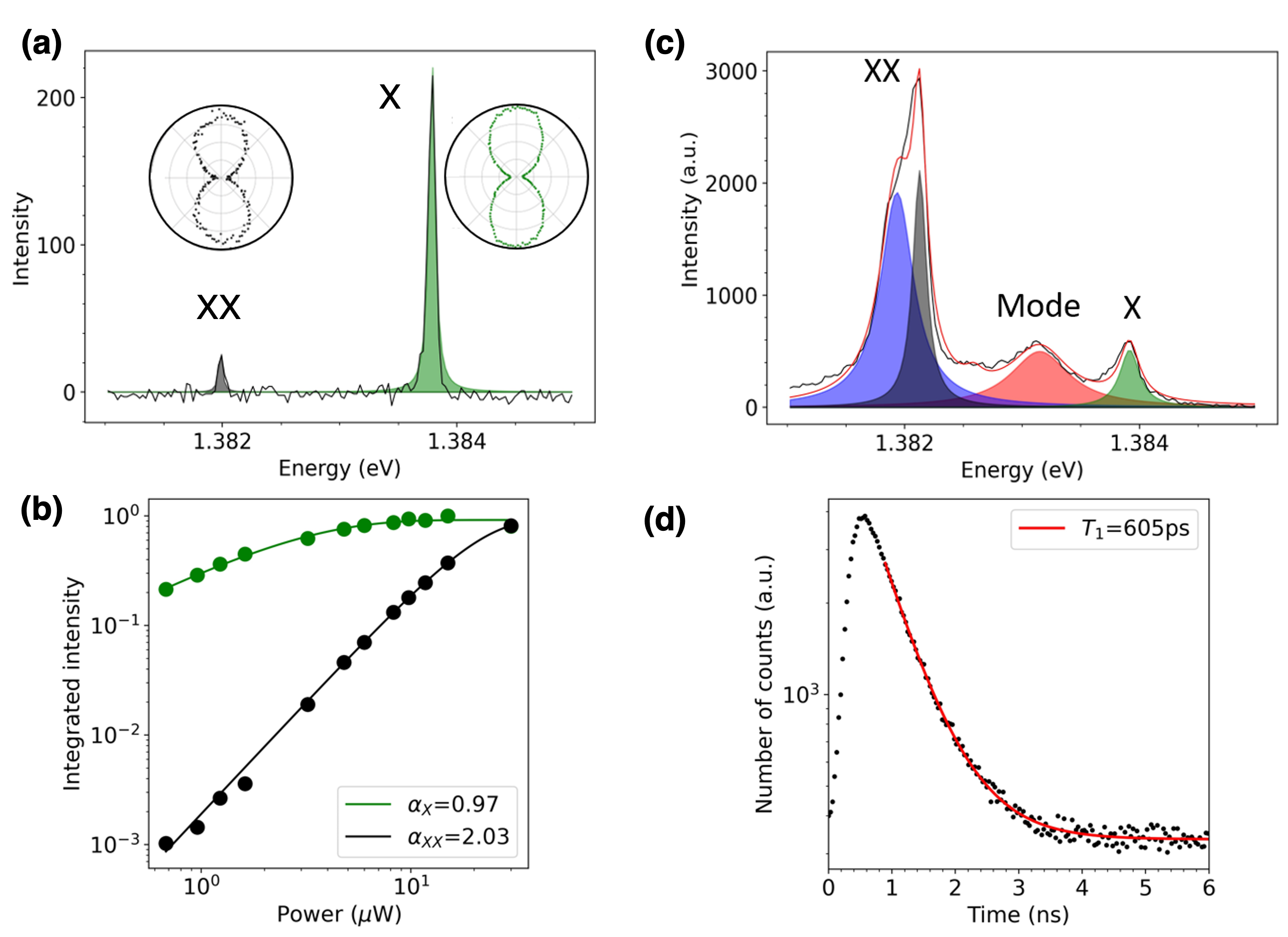}
    \caption{(a) $\mu$-PL spectrum of the cavity-coupled QD under non-resonant pulsed excitation. The excitonic (green) and biexcitonic (black) lines are fitted using Lorentzian functions with respective FWHM $\Gamma_X=38 \mu$eV, $\Gamma_{XX}=63 \mu$eV and energies $E_{XX}=1.3820$ eV, $E_X=1.3839$ eV. Polar diagrams are also shown. (b) Power dependence of the excitonic and biexcitonic $\mu$PL intensities under non-resonant continuous wave excitation. The data (QDs) are fitted using equation $I(P)=I_{sat}(1-\text{exp}(-P/P_{sat}))^\alpha$, with parameters $P^{sat}_X=8\mu W$, $P^{sat}_{XX}=30\mu W$, $\alpha_X=0.97$ and $\alpha_{XX}=2.03$. (c) Micro-PL spectrum under continuous wave excitation at 820 nm with $P\simeq3P^{sat}_X$. The cavity mode (red) is fitted by a Lorentzian function with FWHM $\Gamma_c$=306$\mu$eV and peak energy E$_c$=1.3832eV, corresponding to a quality factor Q=E$_c$/$\Gamma_c$=4520. The emission line of another quantum dot in the vicinity of the cavity (blue) can be observed. (d) Time-resolved micro-PL of the excitonic line under pulsed excitation at 820 nm. The decay of the luminescence is monoexponential : $I(t)=Ae^{-\frac{t}{T_1}}$ with $T_1$ = 605 ps.}
    \label{fig:images_analyse_BQ}
\end{figure*}

We first demonstrate the coupling of the QD to the L3 cavity mode. A non-resonant pulsed Ti:Sa  excitation laser is used in a configuration  where excitation and detection are performed by the top microscope objective.
The µPL spectrum of the QD is presented in Fig. \ref{fig:images_analyse_BQ} a.
We observe two lines attributed to the neutral exciton (green shaded) and biexciton (black shaded). Both lines are vertically polarized parallel to the vertical polarization of the L3 cavity's fundamental mode with a degree of linear polarization of 91\% for the exciton and 90\% for the biexciton. The two transitions are not perfectly polarized because of residual coupling of the QD's emission to the modes above the membrane as will be discussed hereafter. The lines have been attributed to the exciton and the biexciton according to power dependence measurements with continuous wave excitation \cite{babinski_neutral_2008}. As expected, Fig. \ref{fig:images_analyse_BQ}b shows a linear power dependence for the exciton while the biexcitonic line has a superlinear behavior.

The coupling between the QD and the cavity mode can be characterized by the effective Purcell factor F$_P^*$, defined as $F_P^*=T_1^0/T_1$, 
where $T_1^0$ is the exciton's radiative lifetime for a QD located outside the the PhC cavity. Performing time-resolved µPL we measured $T_1^0$ for tens of QDs and took the mean value equal to 
 $T_1^0=1025\pm110\text{ ps}$. 
$T_1$ is the lifetime of the exciton in the cavity-coupled QD and for the QD presented here, it is equal to T$_1$ = 605 ps  (see Fig. \ref{fig:images_analyse_BQ}d).
Thus, F$_P^*\simeq$1.7, which is two orders of magnitude lower than the calculated Purcell factor for this design. 
We attribute this discrepancy to the spatial and spectral mismatch between the QD and the cavity mode. These mismatches are due to the error in the alignment of the cavity to the QD position located by PL mapping, as discussed in Appendix A, and to the small deviation from the targeted cavity parameters during nanofabrication. Indeed, the spectral and spatial matching factors, labelled $\gamma_{\text{spectral}}$ and $\gamma_\text{spatial}$, can be estimated from the definition of the effective Purcell factor \cite{liu_high_2018}  :

\begin{align}
F_P^*&=F_P\times \biggl[
\frac{|\boldsymbol{\mu}\boldsymbol{E_0}|^2}
{|\boldsymbol{\mu}|^2|\boldsymbol{E_0}|^2}
\biggr]
\times\biggl[\frac{1}{1+\bigl[ 2Q(\frac{E}{E_c}-1)\bigr]^2} \biggr] 
\nonumber\\
&=F_P\times\gamma_{\text{spatial}}\times\gamma_{\text{spectral}}
\label{eq:fpetoile}
\end{align}
where $\boldsymbol{\mu}$ is the dipole moment of the QD's transition, $\boldsymbol{E_0}$ is the amplitude of the cavity mode's electric field at the QD's position, $E_c$ is the energy of the cavity mode, Q its quality factor and E is the energy of the optical transition.
By exciting the QD at high pump power (P $\simeq$ 3P$_{sat}$), the cavity mode is fed by phonon-assisted emission \cite{carter_quantum_2013} and can be observed as a Lorentzian line in Fig.\ref{fig:images_analyse_BQ}c, in between the  exciton and biexciton transitions. Another line appears also on the low energy side of the biexciton, likely due to the emission of another QD in the vicinity of the cavity. From this spectrum, we can measure the cavity mode's energy $E_c$ = 1.3832 eV and the quality factor Q = 4520. According to Eq. \ref{eq:fpetoile}, the spectral matching factor between the QD emission at $E = 1.3839$ eV and the PhC cavity mode can be estimated and is $\gamma_\text{spectral}$ = 4.5\% . 
From the values of $F_P^*$, $F_P$ and $\gamma_\text{spectral}$, we obtain the spatial matching factor $\gamma_\text{spatial}$ = 9.1\%. The measured Purcell factor is therefore limited but there is a great potential for improvement if better QD spatial mapping and spectral selection are achieved, as discussed in Appendix \ref{app:mapping}.

\begin{figure}
    \centering
    \includegraphics[width=0.45\textwidth]{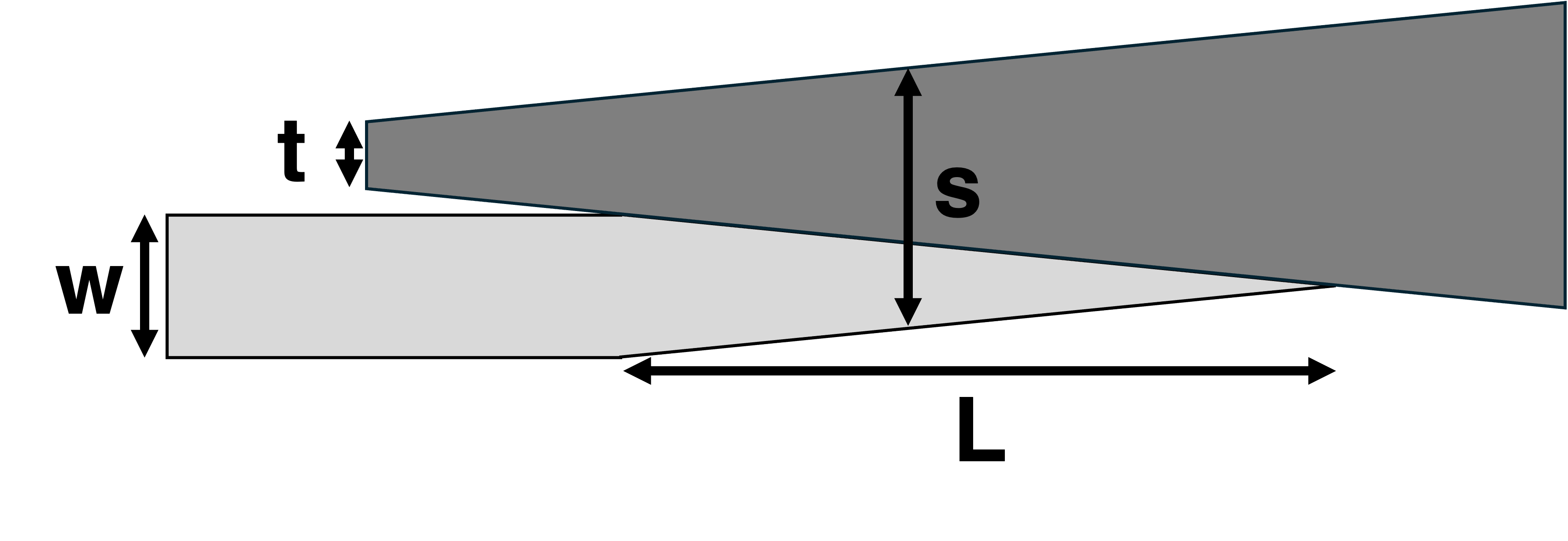}
    \caption{Schematic of the relevant parameters for the simulation of the adiabatic coupling between the fiber (dark grey) with a tip diameter t and the waveguide (light grey). The waveguide's width decreases from w to 0 over the coupling length L. Over this coupling length, the width of the "superguide" which arises from the two materials is assumed to be constant and equal to s.}
    \label{fig:schem_couplage}
\end{figure}

\begin{figure}
    \centering
    \includegraphics[width=0.45\textwidth]{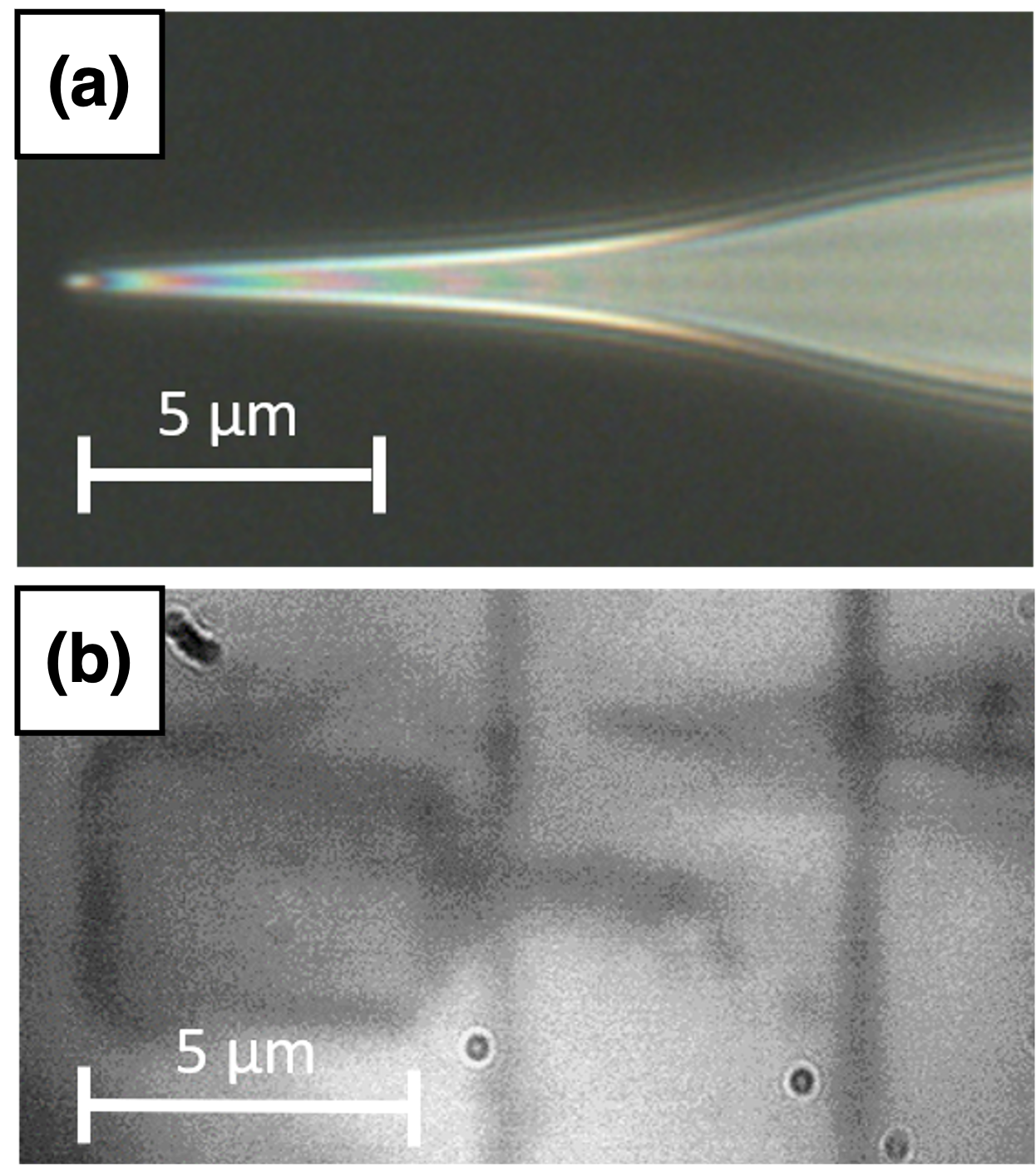}
    \caption{(a) Optical microscope image of the realized stretched tapered fiber. (b) Optical microscope image of the sample (down left) and the tapered fiber (up right) approaching the waveguide inside the cryostat. The vertical lines seen in the image correspond to the edges of the membrane and the sample respectively, as depicted Fig. \ref{fig:image_MEB_II}(b).}
    \label{fig:images_MEB_fibre}
\end{figure}

The efficiency of the collection by the fiber depends on the adiabatic coupling between the tapered waveguide and the fiber.
The relevant parameters for the simulation of the adiabatic coupling are presented Fig. \ref{fig:schem_couplage}. 
The coupling between the waveguide with refractive index 3.45 (light gray) and the fiber with refractive index 1.46 (dark gray) arises from the adiabatic transfer of the mode from the waveguide of width w to the tapered fiber over the coupling length L. The adiabatically coupled tapered GaAs waveguide and tapered fiber can be considered as a single "superguide" with a constant width s.
The fiber is in contact with the waveguide, and any small misalignement between the fiber's axis and the waveguide's axis is compensated by the slight flexibility of the fiber.
The fiber can also be moved along the coupling axis, allowing us to tune the "superguide" size s, whose minimum value is limited by the diameter of the fiber tip ($t\simeq350$ nm, see Fig. \ref{fig:images_MEB_fibre} a) and maximum value by how close the fiber can be approached to the sample.
This can be seen on Fig. \ref{fig:images_MEB_fibre} b, which shows an optical image taken during the experiment at 4K when moving the fiber with the piezoelectric actuators to control its position with respect to the waveguide. The sub-micrometer mechanical vibrations are optically magnified giving the impression of a blurry image.

\begin{figure}
    \centering
    \includegraphics[width=0.45\textwidth]{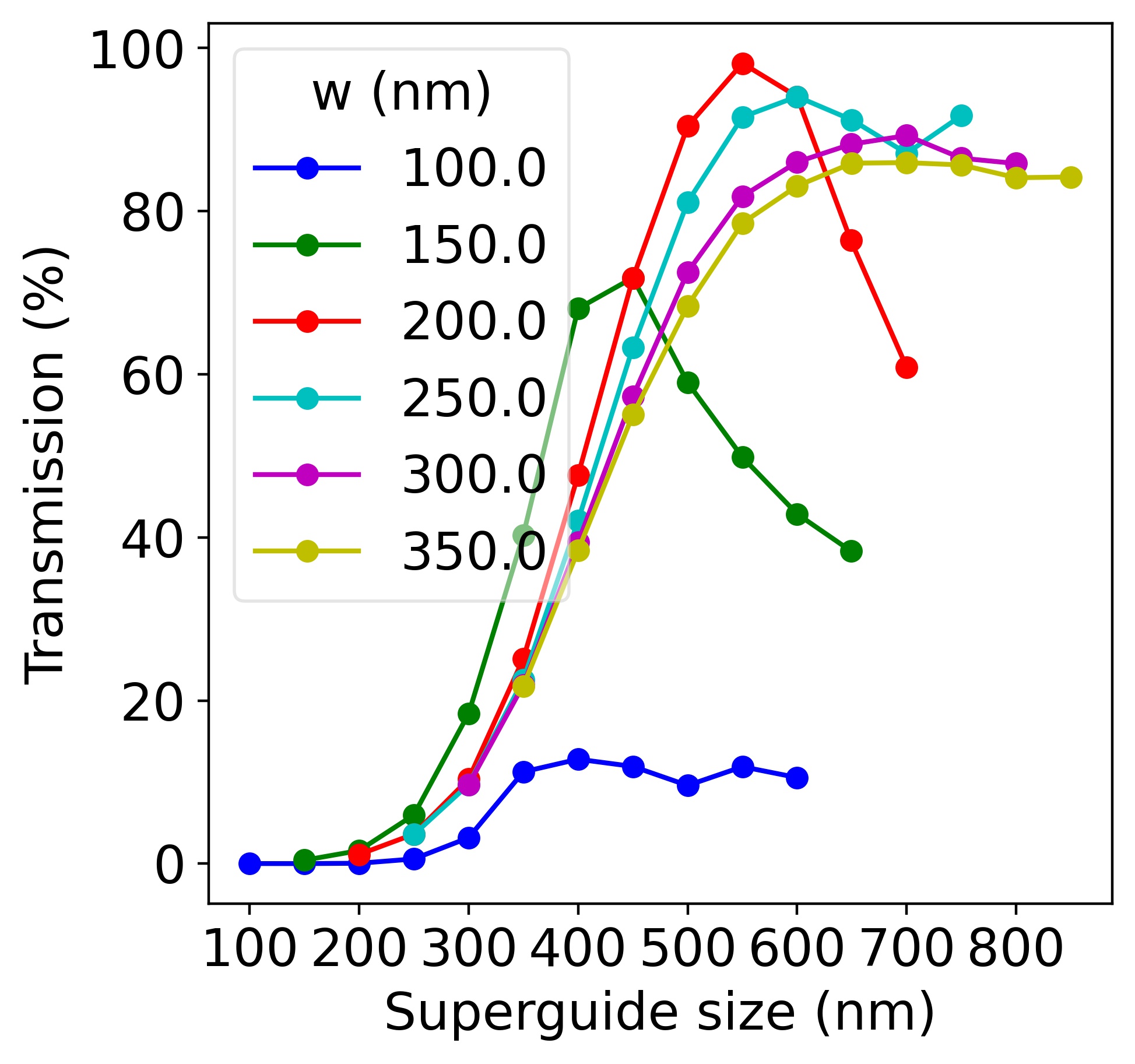}
    \caption{Simulated variation of the adiabatic coupling efficiency as a function of s and w for L = 5000 nm for a fiber with a negligible tip radius. For w = 200 nm and s = 550 nm, the coupling is over 99 \%.}
    \label{fig:wandsvariation}
\end{figure}

FDTD simulations (not shown here) were performed for various values of w, s and L. 
The minimum coupling length for which the coupling is over 99\% is L = 5000 nm. The variation of the calculated coupling efficiency as a function of s for various w is presented in Fig. \ref{fig:wandsvariation}.
It can be seen that the coupling is maximized for w=200 nm and s=550 nm. 

\begin{figure}
    \centering
    \includegraphics[width=0.45\textwidth]{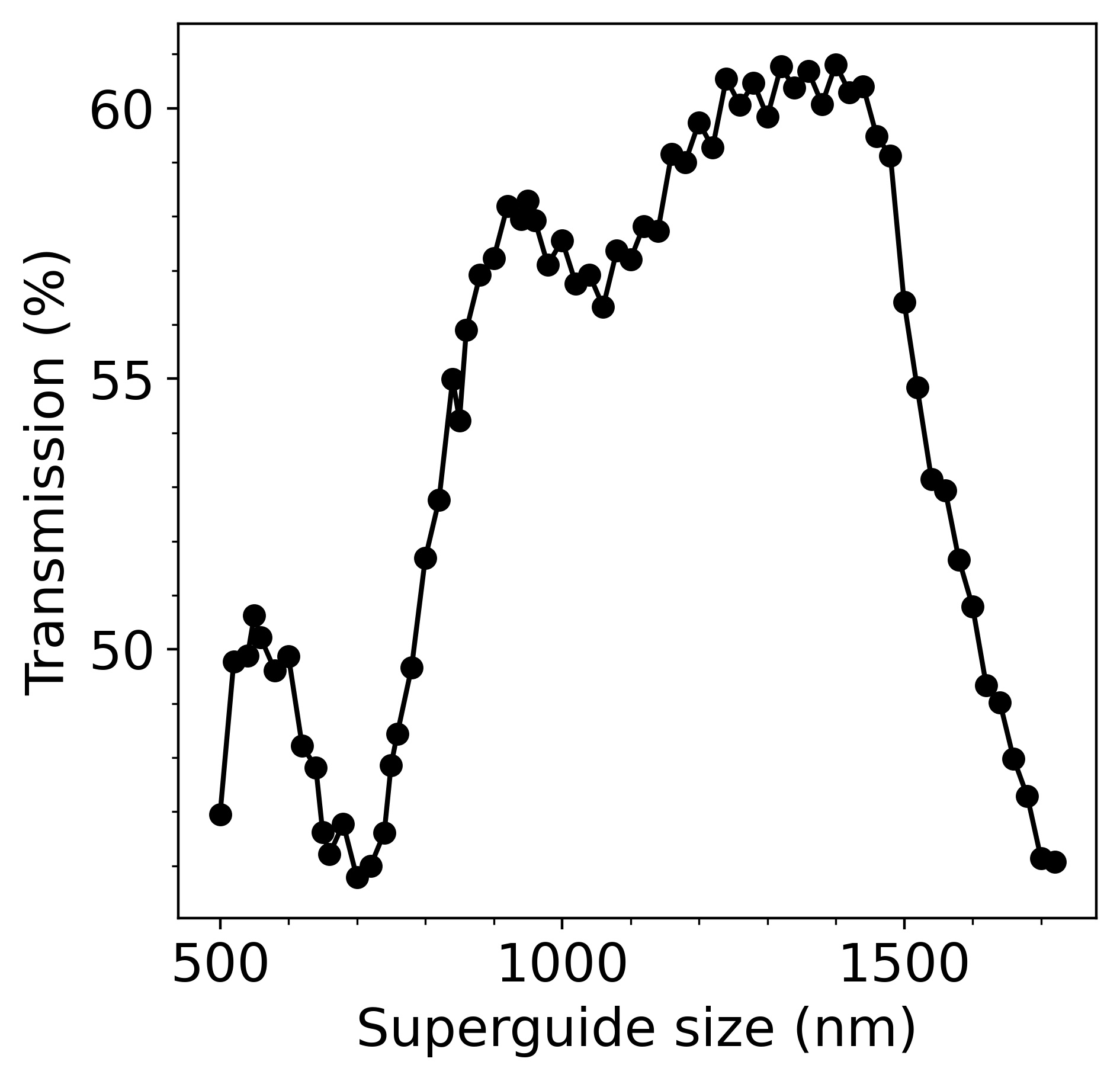}
    \caption{Simulated variation of the adiabatic coupling efficiency as a function of the superguide size for a coupling length L = 2500 nm and a waveguide width w = 500 nm. The maximum transmission is obtained for a superguide size of 1400 nm and is T = 60.8\%.}
    \label{fig:s_var}
\end{figure}

In order to ensure the mechanical stability of the free-standing waveguide, the coupling length L should be as small as possible, so the tapered waveguide was fabricated with a base width of 500 nm and a length of 2.5$\mu$m.
The resulting simulated transmission for these dimensions as a function of the supermode's size s is plotted Fig. \ref{fig:s_var}.
With the current experimental setup, it is possible to tune the superguide size s by moving the fiber along the waveguide axis in order to maximize the coupling and reach 61 \%. Thus, we will use the value $\eta_{WG/fiber}$ = 0.61  to compute the theoretical transmission efficiencies in the following.

The fiber transmission efficiency  $\eta_{fiber}$ is equal to :
\begin{align}
\eta_{fiber}=&\eta_{cav/PCWG}\times\eta_{PCWG/WG}\times\eta_{holder}\nonumber
\\
&\times\eta_{WG/fiber}\times\eta_{fiber/fiber}
\end{align}
where all the values have been calculated using FDTD simulations, except for the transmission of the bonded fibers which was measured as $\eta_{fiber/fiber}=0.998$. The calculation of the coupling between the cavity and the PCWG $\eta_{cav/PCWG}=0.88$ is presented in Appendix \ref{app:pcparam}. The coupling between the PCWG and the GaAs bulk waveguide $\eta_{PCWG/WG}=0.9$ is not exactly unity because of the difference in phase velocity between the PCWG and the bulk waveguide, leading to reflection at the PhC output \cite{miyai_analysis_2002, miyai_structural_2004,sanchis_analysis_2004}. Finally, $\eta_{holder}=0.985$ takes into account the free-standing waveguide losses due to the tether points located $\simeq$1$\mu$m after the PCWG transition to a bulk free-standing waveguide. By replacing all the parameters by their simulated values, we obtain $\eta_{fiber}=0.46$.
The intrinsic losses of the PCWG are considered negligible compared to other losses due to the short length of the PCWG.

The fraction of the cavity's mode field intensity collected by an objective of N.A 0.65 located above the membrane is :
\begin{equation}
    \eta_{objective}=(1-\eta_{cav/PCWG})\times\eta_{0.65}\times\eta_{optical path}
\end{equation}
where $\eta_{0.65}$ is the collection efficiency of the mode by the top microscope objective, as explained in Appendix \ref{app:pcparam}. 
$\eta_{optical path}$ is the collection efficiency of our optical setup before the spectrometer, which takes into account the losses of the confocal microscopy setup due to optical components and was measured to be $\eta_{optical path}$= 0.39, leading to $\eta_{objective}=0.044$. 
From these two values we can compute the expected ratio of the intensities collected by the fiber and by the objective, $\zeta=10.45$.

\begin{figure}
    \centering
    \includegraphics[width=0.45\textwidth]{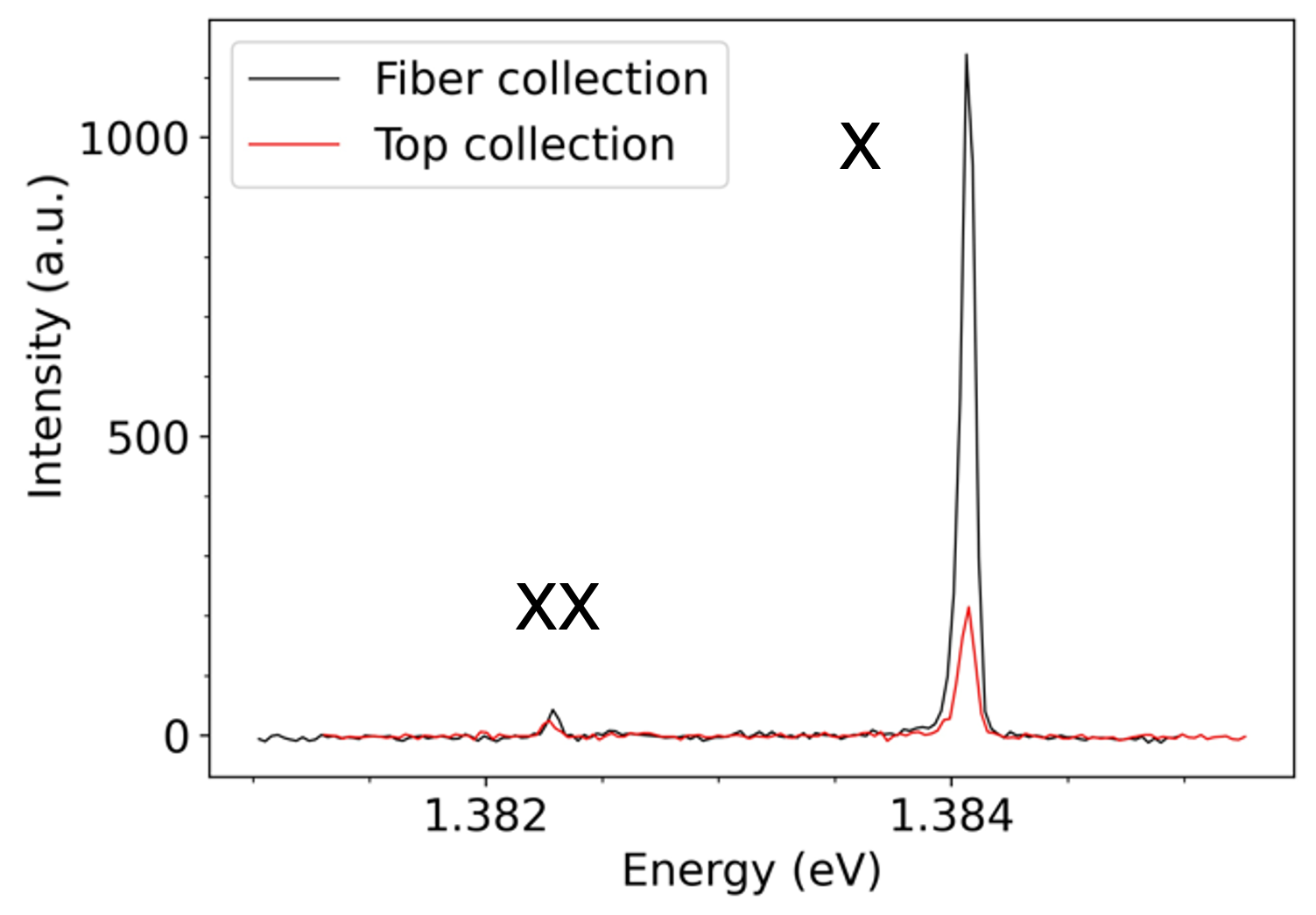}
    \caption{ µPL spectrum of the quantum under top excitation using fiber (black curve) and top (black curve) collection. The intensity of the exciton's PL collected by the fiber path is 4.78 times more than that collected by the objective path. A much smaller enhancement of the biexciton's PL collection is observed, since it is not as well matched to the wavelength for which the photonic nanostructure is optimized.}
    \label{fig:PL_fiber}
\end{figure}

The adiabatic coupling can be experimentally characterized by estimating the efficiency ratio $\zeta$ when comparing the intensities of the luminescence collected by the objective and by the fiber (see Fig. \ref{fig:PL_fiber}).
We obtain $\zeta_{exp} \simeq5$, which is less than the expected factor $\zeta$ = 10.45. 
We attribute this discrepancy to the non-negligible coupling of the QD's emission to the modes of the electromagnetic field above the slab due to the non-perfect QD-cavity coupling. 
Indeed, the computation of $\zeta$, detailed in Appendix B, assumes that all the photons emitted by the QD are emitted in the cavity mode, which is not the case. 
This also explains the fact that the luminescence of the exciton is not fully vertically polarized as previously noticed. 
Additional losses may also occur due to imperfections in the processing of the photonic nanostructure, leading to variation of the holes radii.

\begin{figure}
    \centering
    \includegraphics[width=0.45\textwidth]{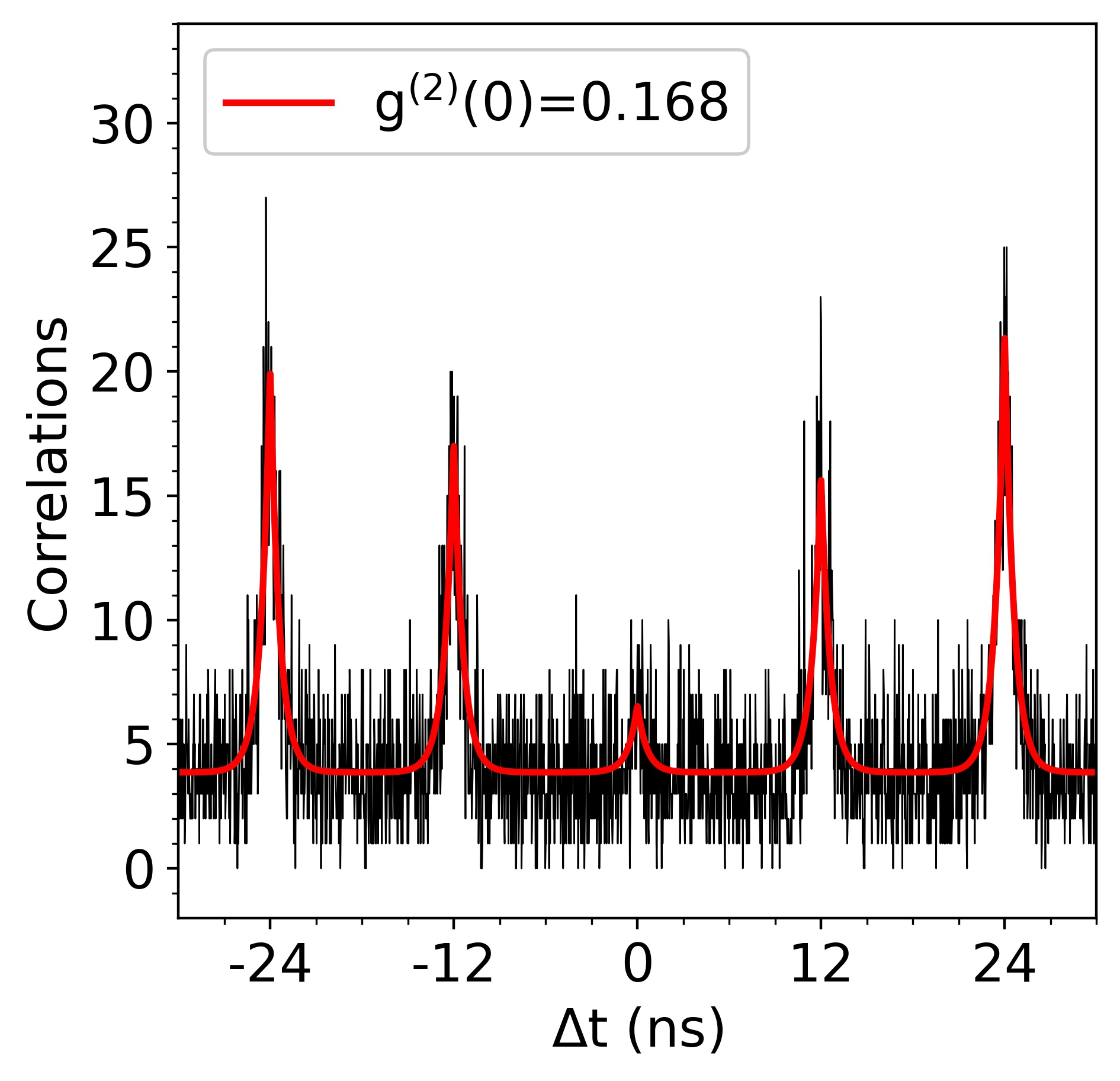}
    \caption{ (a) Second order correlation measurements under pulsed non-resonant excitation at pump power close to saturation.
    The data are fitted using the function $f(t)=A_0 \text{exp}(-|t|/T_1)+\sum_{i\neq0}A_i \text{exp}(-|t-n_i t_0|/T_1)$ where the index i runs over all the peaks. The second order correlation function at zero delay is equal to $g^{(2)}(0)=A_0/<A_i>_{i\neq0}$.}
    \label{fig:g2}
\end{figure}

The purity of single photon emission collected by the fiber can be characterized in a Hanbury Brown Twiss experiment. The QD is excited using pulsed excitation at 820 nm and an additional very low power He-Ne laser is used to stabilize the electrostatic environment of the QD \cite{nguyen_optically_2012}. 
Fig. \ref{fig:g2} shows the second order correlation function which at zero-delay is $ g^{(2)}(0)=0.168$. 
The remaining correlations at zero delay are likely due to multiphoton emission due to the high off-resonant pump power  ($P\,\simeq P_{sat}$) which was used in this experiment. 

Performing photon correlation measurements under resonant excitation will enhance the purity of the single photon source and is the next step to achieve in order to demonstrate bright single photon emission coupled to off-chip optical fibers.
To achieve this, the PCWG spatial length needs to be increased so that the spatial seperation between excitation and collection positions is great enough to avoid scattered laser light being collected by the fiber.


\section{Conclusion}

We demonstrated  the in-plane collection of single photons emitted by a QD embedded in a PhC cavity through adiabatic coupling with a tapered fiber. 
This was achieved using a PCWG coupled to a PhC cavity to guide the single photons into a tapered bulk GaAs waveguide which was then adiabatically coupled to the optical fiber.
The collection efficiency could be experimentally estimated by comparing the intensities of the luminescence collected by the top microscope objective and by the tapered fiber, taking into account the coupling of the L3 cavity to the PCWG and the losses of the objective and fiber collection paths. We obtain a ratio of 5, only a factor of 2 less than the collection efficiency predicted by simulations.
Under non-resonant excitation, we measured the value of the second order correlation function at zero delay g$^{(2)}$(0)=0.17.



Further improvements in this photonic nanostructure can be expected by achieving better spatial and spectral matching of the emitter with the photonic cavity. This will increase the fraction of emitted photons that are coupled into the photonic crystal cavity mode and subsequently funneled into the PCWG.
Finally, a near-unity theoretical adiabatic coupling efficiency between the tapered GaAs waveguide and a tapered optical fiber is predicted if the coupling length between the waveguide and the tapered fiber is increased from the 2.5 microns used here to 5 microns. Such a high coupling efficiency to an optical fiber is highly attractive for the realization of fiber-coupled single photon sources for practical applications.

\section*{Acknowledgments}

This work was funded by the French National Research Agency (ANR “ISQUAD” Grant No. ANR-18-CE47-0006-01), by the Paris Île-de-France Région in the framework of DIM SIRTEQ and by the Cluster of Excellence MATISSE led by Sorbonne Université. A.B. and B.E. are grateful to the Quantum Information Center Sorbonne (QICS) for specific fundings. 

\appendix 

\section{Mapping of the photoluminescence}
\label{app:mapping}

The probability of a QD being located at the center of a PhC cavity is very low due to the stochastic formation of QDs during their growth.
Moreover, the emission wavelengths of the QDs are distributed over a broad spectral range, thus reducing the probability of spectral matching.
To enhance the spatial and spectral matching factors, a µPL mapping is performed in order to determine the positions of the QDs relative to Ti/Au markers before etching the photonic nanostructures.
The QDs are excited using a green LED (535 nm) and the markers are illuminated using an infrared LED (910 nm). The light scattered by the sample is then spectrally filtered to suppress the reflection in the green and select the QDs emitting within the narrow spectral window that we choose.

\begin{figure}
    \centering
    \includegraphics[width=0.45\textwidth]{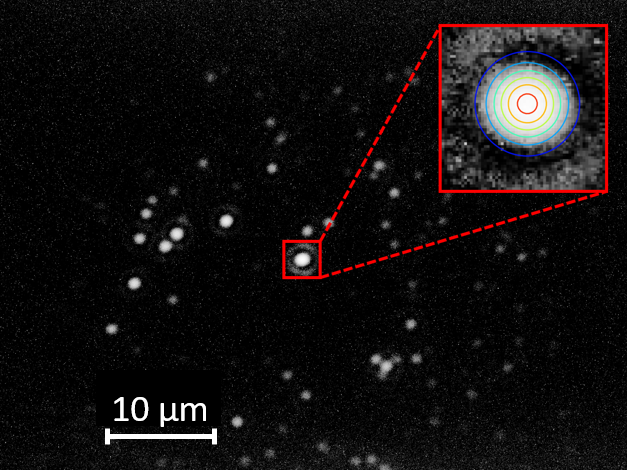}
    \caption{µPL imaging of the QDs luminescence. The inset shows the intensity contours of the 2D Gaussian fit to the brightest QD of the image. The Gaussian fit allows the dot center to be found with sub-pixel accuracy.}
    \label{fig:maping_QD}
\end{figure}

The obtained diffraction-limited image of the QDs' PL is presented Fig. \ref{fig:maping_QD}. 
The QDs' positions are found with sub-pixel accuracy (1 pixel = 50 nm) using a 2D Gaussian fit, whose intensity contour lines are plotted in the inset of Fig. \ref{fig:maping_QD}.
We estimate the error on the QDs positions of the order of 25 nm.

\begin{figure}
    \centering
    \includegraphics[width=0.45\textwidth]{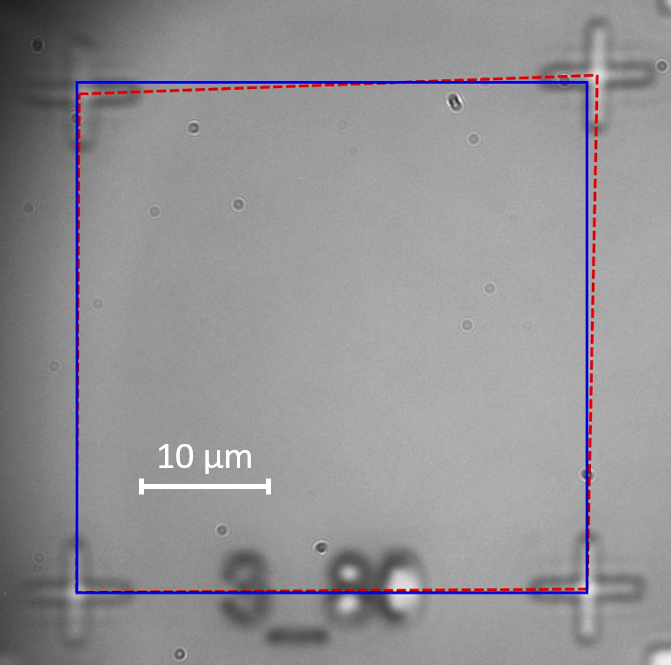}
    \caption{Deviation of the polygon formed by the center of the 4 crosses (dashed red line) from an ideal square (blue line).}
    \label{fig:mapping_cross}
\end{figure}

The imaging of the markers is presented in Fig. \ref{fig:mapping_cross}.
The dotted red line defines the polygon shaped by the center of the 4 crosses defining a zone. As it can be seen, this polygon deviates from an ideal square, plotted in blue, due to image distortion, defined as :
\begin{equation}
    D=\sqrt{\langle \frac{|\Delta x_{i}|}{L}\rangle^2+\langle\frac{|\Delta y_{i}|}{L}\rangle^2}
\end{equation}
where $\Delta x_{i}=x_{\text{cross},i}-x_{\text{square},i}$ and $\Delta y_{i}=y_{\text{cross},i}-y_{\text{square},i}$, with the index i running over the four polygon corners and L being the side length of the square, defined as the mean value of the distances between neighboring crosses.
The resulting distortion is of the order of 1\%, equivalent to 400 nm for a square of 40 $\mu$m x 40 $\mu$m. To reduce the impact of the distortion on the QD's positioning relative to the markers, the QD's position is taken relative to the position of the nearest cross, so that the distortion error doesn't propagate over a  too large distance.
We estimate the resulting total error on the QDs positions to be better than $\pm$200 nm.
This error could be reduced by a factor 4 using grid markers \cite{liu_cryogenic_2017} combined with distortion correction algorithms.

The probability of spatial matching between a quantum and a photonic nanostructure is therefore greatly enhanced in comparison to a sample without spatial mapping of the PL, as we almost always observe QD's PL near the PhC cavities. 
The spectral matching depends on the bandwidth of the spectral filters used for the mapping and the uncertainty on the radii of the holes of the photonic nanostructure.
For the sample presented here, the spectral window which was used was 910 $\pm$ 10 nm, but using combinations of tunable high- and low-pass filters it can be reduced to $\pm$2 nm at the desired central wavelength.
The uncertainty on the central wavelength of the cavity mode, which was measured to be $\pm$7 nm, would then be the only limiting factor for spectral matching without requiring any additional in-situ spectral tuning.

\section{Photonic nanostructure parameters}
\label{app:pcparam}

All the parameters of the designed photonic nanostructures were optimized using the FDTD simulation software provided by Lumerical®. 
We give in Table 1, all the relevant parameters of the photonic nanostructures. 
\begin{table}
\caption{Photonic nanostructure parameters. The first column gives the nanostructure parameters for a cavity (see Fig. \ref{fig:scheme_coupling}) with a simulated central wavelength at 910 nm. The second column gives the estimated values for the same design but with the experimentally observed central wavelength of 897 nm.}
\begin{ruledtabular}
\begin{tabular}{ccc}
Parameter&Simulated value&Value\\
(notation)&(nm)&(nm)\\
\hline
Distance to buffer (d) & 500&500 \\
Lattice constant (a)& 231&231\\
Holes radii (r ) & 70&74\\
First neighbors radii (r$_1$) & 70&74\\
First neighbors shift (|$\Delta$x|) & 40&40\\
Second neighbors radii (r$_2$) & 74&78\\
W1 waveguide first neighbors radii (r$_1^{wg}$) & 64&68\\
Bulk waveguide width (w) & 500 &500\\
Coupling length (L) & 1500 &1500\\
\end{tabular}
\end{ruledtabular}
\label{table:param}
\end{table}

The structure is originally designed for a QD emitting at 910 nm, but due to the uncertainties on the holes radii, the mode of the cavity was at 897 nm.
In order to estimate the various properties of the structure, we assumed that the holes radii were all increased by the same amount.
By running FDTD simulations and varying this increase in hole size, we obtain a mode centered at 897 nm for an increase in hole radius of 4 nm over the whole pattern. 
These larger hole sizes were used to compute the values of the collection efficiencies.

\begin{figure}
    \centering
    \includegraphics[width=0.45\textwidth]{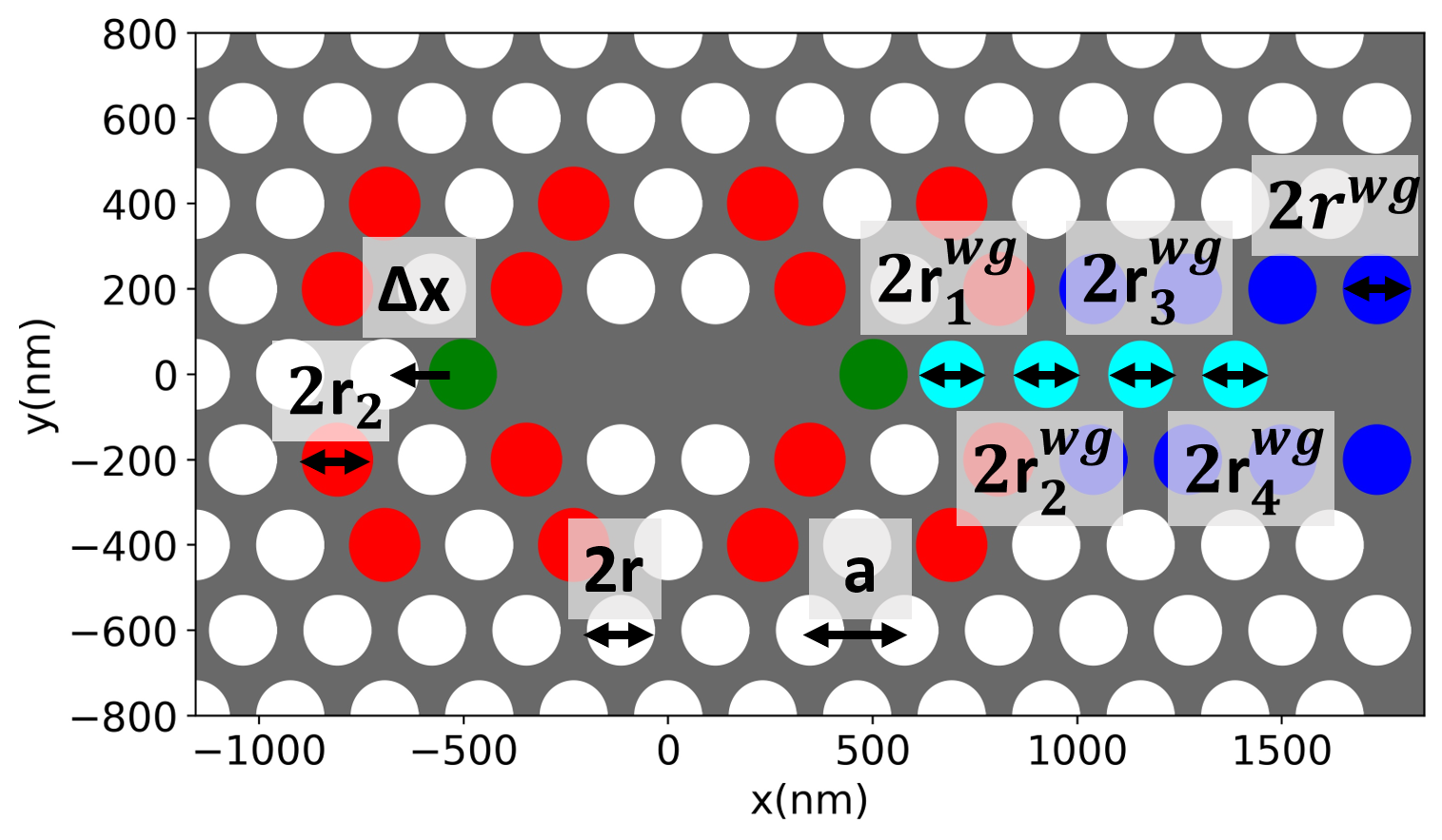}
    \caption{Relevant parameters for the coupling between the L3 cavity and the PCWG. The lattice's parameters are a and r. The green holes are shifted outwards of $\Delta x$ to enhance the quality factor, while the red holes radii r$_2$ is modified to enhance the collection efficiency by an objective located above the membrane. The light blue holes radii are modified to tune the coupling between the cavity and the PCWG. The dark blue holes are modified to minimize the PCWG losses.}
    \label{fig:scheme_coupling}
\end{figure}

A schematic showing the relevant parameters for optimizing the coupling between the cavity and the PCWG is presented Fig. \ref{fig:scheme_coupling}.
The lattice parameters a and r are chosen such that the wavelength of the QD lies within the PhC band gap. 
Without any modification of the holes in the vicinity of the cavity, the fundamental mode of the L3 cavity has a quality factor Q $\simeq$ 5000, and its far field emission profile is vertically polarized \cite{akahane_high-q_2003, akahane_fine-tuned_2005}, such that the collection efficiency of the light radiated by the cavity above the membrane and collected by a microscope objective is $\eta_\text{0.65}\simeq0.55$ (in the following, all the values are given for an objective of NA 0.65).
By shifting the holes depicted in green Fig. \ref{fig:scheme_coupling} by |$\Delta$x|, the quality factor of the cavity Q can be increased \cite{srinivasan_momentum_2002}, and by modifying the radii r$_2$ of the holes depicted in red, $\eta_\text{0.65}$ can be enhanced \cite{kim_vertical_2006, hagemeier_h1_2012, portalupi_planar_2010, tran_directive_2009}. The output of these two processes gives $Q_{\text{without WG}}=27 000$ and $\eta_\text{0.65}=0.71$.

\begin{figure}
    \centering
    \includegraphics[width=0.45\textwidth]{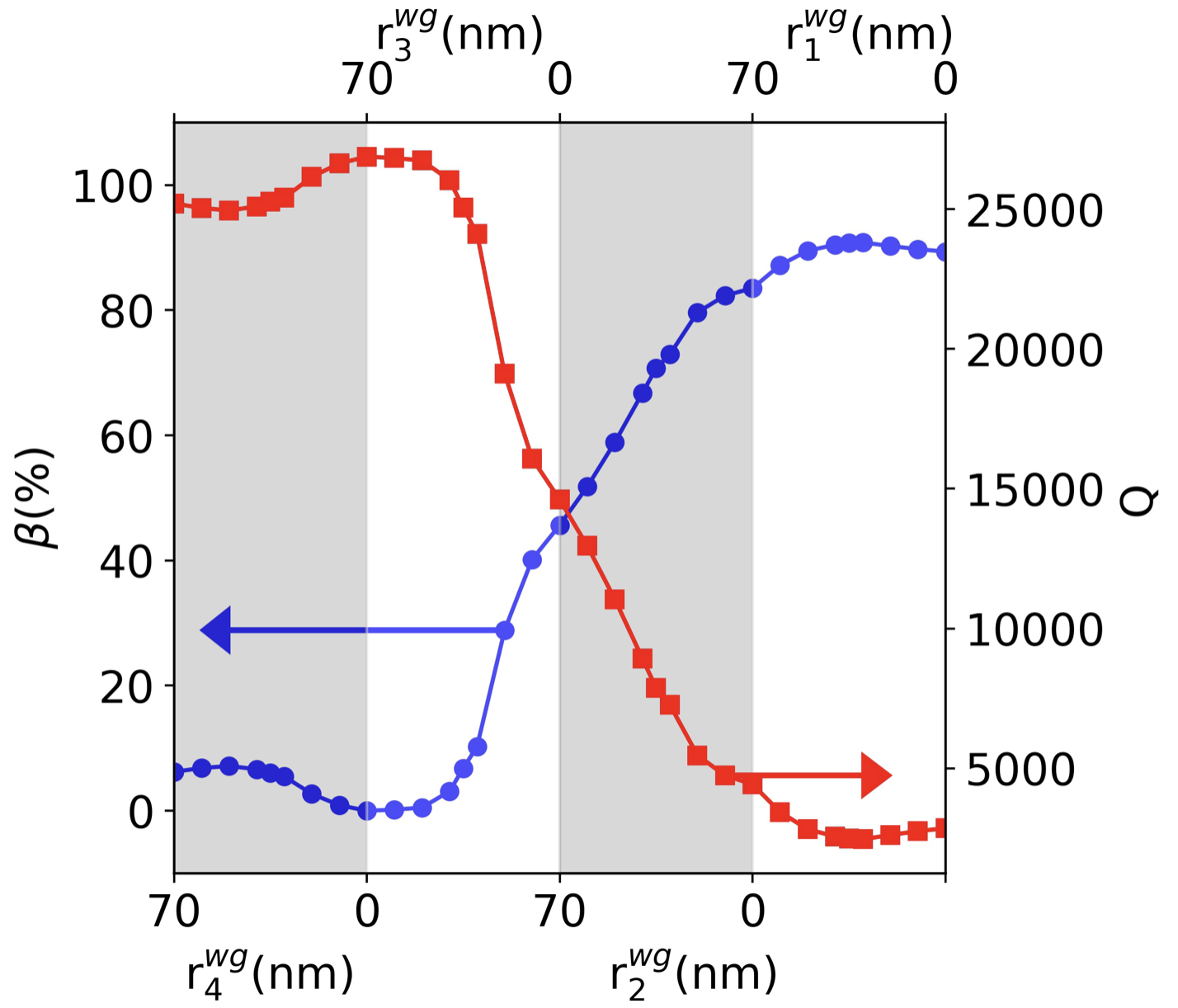}
    \caption{Simulated variation of the coupling of the L3 cavity to the PCWG as a function of the radii of the four holes neighboring the cavity. In each of the four regions of the graph the radius of one of the holes decreases from 70nm to 0nm, while the radii of the other holes remain the same. The radii of r$_1^{wg}$/r$_2^{wg}$/r$_3^{wg}$/r$_4^{wg}$  are for each region, from left to right, in nanometers : 70$\rightarrow$0/70/70/70, 0/70$\rightarrow$0/70/70, 0/0/70$\rightarrow$0/70 and 0/0/0/70$\rightarrow$0.}
    \label{fig:var_rwg}
\end{figure}

The PCWG's first neighbors radii, shown in dark blue in Fig. \ref{fig:scheme_coupling}, are modified to increase the group velocity of the photons traveling at the exciton's wavelength  \cite{chen_creating_2008}, thus reducing the propagation losses \cite{tanaka_group_2004,krauss_slow_2007,wasley_disorder-limited_2012,waks_coupled_2005}. 
Due to the spatial overlap between the cavity mode and the PCWG's fundamental mode, the energy of the cavity leaks into the PCWG.
The coupling between the cavity and the PCWG is then tuned by modifying the holes radii\cite{okano_coupling_2003,kim_coupling_2004,coles_waveguide-coupled_2014} r$_1^{\text{wg}}$, r$_2^{\text{wg}}$, r$_3^{\text{wg}}$ and r$_4^{\text{wg}}$, shown in light blue Fig.\ref{fig:scheme_coupling}.
The coupling is then equal to 
\begin{equation}
\eta_\text{cav/PCWG}=1-\frac{Q_{\text{with WG}}}{Q_{\text{without WG}}}
\end{equation}
The relevant parameters are presented Fig. \ref{fig:scheme_coupling} and the variation of the coupling with the radii of the holes between the cavity and the PCWG is presented Fig. \ref{fig:var_rwg}.
The coupling can be tuned from 0\% to 91\%. For this sample, we chose r$_1^{wg}$=70 nm and r$_2^{wg}$=r$_3^{wg}$=r$_4^{wg}$=0, such that $\eta_\text{cav/PCWG}=0.88$.

\newpage
\end{document}